\documentclass[]{tPHM2e}

\usepackage[utf8]{inputenc}
\usepackage{subfigure}
\usepackage{graphicx}
\usepackage{amsmath,amssymb}
\usepackage{xspace}

\begin{document}

\newcommand{\etal}{\textit{et al}.\@\xspace}
\newcommand{\ie}{\textit{i.e.}\@\xspace}
\newcommand{\cf}{\textit{cf}.\@\xspace}

\newcommand{\mi}{\mathrm{i}}
\newcommand{\md}{\mathrm{d}}


\doi{10.1080/14786430902976794}
\issn{1478-6443}
\issnp{1478-6435}
\jvol{89} \jnum{19} \jyear{2009} \jmonth{1 July}
\markboth{E. Clouet}{Philosophical Magazine}

\title{Elastic energy of a straight dislocation \\
and contribution from core tractions}
\author{Emmanuel Clouet$^{\dagger}$\thanks{$^{\dagger}$ Email: emmanuel.clouet@cea.fr}
\\\vspace{6pt}
CEA, DEN, Service de Recherches de Métallurgie Physique, \\
F-91191 Gif-sur-Yvette, France
\\\vspace{6pt}
\received{\today}
}

\maketitle
\begin{abstract}
	We derive an expression of the core traction contribution
	to the dislocation elastic energy within linear anisotropic
	elasticity theory using the sextic formalism. 
	With this contribution, the elastic energy
	is a state variable consistent with the work 
	of the Peach-Koehler forces.
	This contribution needs also to be considered when extracting
	from atomic simulations core energies. 
	The core energies thus obtained are real intrinsic dislocation properties:
	they do not depend on the presence and position of other defects.
	This is illustrated by calculating core energies 
	of edge dislocation in bcc iron, where we show that
	dislocations gliding in \{110\} planes
	are more stable than the ones gliding in \{112\}
	planes.

	\bigskip

\end{abstract}

\section{Introduction}

One important quantity controlling the physics of dislocations is
their elastic energy. 
It is defined as the integral of the elastic energy density
over the whole volume except a small core region surrounding the dislocation line.
This excludes the region around the dislocation
core where elasticity does not apply 
because of the too high strains. 
Using Gauss theorem, the elastic energy can be decomposed in
two contributions:
\begin{itemize}
	\item the one corresponding to an integration
		along the dislocation cut of the work necessary 
		to create the dislocation. 
		As it is well known, this contribution varies with
		the logarithm of a characteristic distance 
		of the dislocation microstructure. 
	\item the contribution arising from the work 
		done by the tractions exerted on the tube
		which isolates the dislocation core.
		The corresponding contribution to the elastic energy
		is known as the contribution of the core tractions.
\end{itemize}
This last contribution is sometimes forgotten. Indeed, the cut contribution
is usually the dominant one.
Moreover, the core traction contribution disappears when one tries to define
the elastic energy of an isolated infinite straight dislocation 
because of the external cylinder that has to be introduced
to prevent the elastic energy from diverging.
Nevertheless, if one wants the dislocation elastic energy
to be a state variable, \ie a variable that only depends on the 
current state and not on the transformation path used to reach this state,
both contributions need to be taken into account. 
Bullough and Foreman \cite{BUL64} already showed that the elastic energy 
of a dislocation loop does not depend on the hypothetical creation mechanism
only when both contributions are considered. 
Lothe and Hirth \cite{LOT05} also noticed that the definition 
of the elastic energy of a straight dislocation could not be consistent 
with the work of the Peach-Koehler forces if the contribution 
of the core tractions was not included in the elastic energy.
Finally, Gavazza and Barnett \cite{GAV76} showed that 
the dislocation elastic energy part associated with the core tractions
leads to a contribution to the self-force acting on a dislocation loop.

Despite its importance, no analytical expressions of the core traction contribution 
to the elastic energy seems to exist in the literature when the elastic anisotropy 
is considered.
The anisotropic linear elasticity theory of dislocations has been developed
in the past sixty years. 
Eshelby, Read and Shockley \cite{ESH53} were the first to express 
the anisotropic elastic field created by a straight infinite dislocation.
Their formalism was then enriched by Stroh \cite{STR58,STR62}, 
leading to what is known as the sextic formalism.
Latter, it was shown that the elastic field of a dislocation 
of arbitrary shape can be obtained from the fields of straight
infinite dislocations\footnote{For a review, see Ref. \cite{BAC80}.}.
But no expression of the contribution of the core tractions 
to the elastic energy has been obtained until now.
The variation of this contribution with a rotation 
of the dislocation cut is known \cite{CAI03,LI04}.
It actually corresponds to the angular dependence 
of the interaction energy between two dislocations 
derived by Stroh \cite{STR58}.
But an expression of the absolute value of this 
contribution is not available.
Such an analytical expression exists under the assumption 
that the elastic constants are isotropic \cite{BUL64}. 
The few studies that considered the elastic anisotropy as well 
as the contribution of the core tractions \cite{BAC70,GOS94,GOS96,DUD08} 
calculated this contribution
with a numerical integration along the surface of the core cylinder. 
It is the purpose of this article to obtain an analytical 
expression of this contribution in the framework of the sextic formalism.
Such an expression could be used then in further studies.
Potential applications are the extraction from atomic simulations
of dislocation energy properties like core energies 
\cite{XU96,ISM00,BLA00,CAI01,CAI03,LI04,CLO09},
calculations of dislocation loop self energy \cite{BUL64,BAC70,DUD08}, 
or computation of the self-force acting on a dislocation loop \cite{GAV76}.
Such an application is presented in this article where core energies
of edge dislocation in $\alpha$-iron are determined.

In the first section, we define the elastic energy of a straight 
dislocation so as to clearly make appear the contribution 
of the core tractions. 
We then use isotropic elasticity to illustrate this definition
and highlight the importance of this contribution for the 
coherency of the elastic energy definition.
The analytical expression of the core traction contribution
within anisotropic linear elasticity is obtained in the following
section using the sextic formalism.
We finally illustrate the consistency of our results by studying 
edge dislocations in $\alpha$-iron and calculating their 
core energies with different simulation methods.

\section{Elastic energy of a straight dislocation}
\label{sec:energy}

We assume in this article that the elastic field created by a dislocation
can be reduced to the Volterra solution. 
Eshelby \etal \cite{ESH53} have indeed shown that a straight dislocation 
in an infinite elastic medium creates in a point defined by its cylindrical 
coordinates $r$ and $\theta$ a displacement given by the Laurent series 
\begin{equation*}
	{\bm u}(r,\theta) = {\bm v}\ln{(r)} + {\bm u}_0(\theta)
	+ \sum_{n=1}^{\infty}{ {\bm u}_n(\theta) \frac{1}{r^n} }.
\end{equation*}
The two first terms of this series, ${\bm v}\ln{(r)} + {\bm u}_0(\theta)$
correspond to the Volterra solution: this describes the elastic field
of the dislocation far enough from its core. 
The remaining terms of the series ($n\geq1$) are the dislocation core field \cite{TEO82}, 
which arises from non-linearities in the crystal elastic behavior
and from perturbations due to the atomic nature of the core.
We do not consider this part of the elastic field 
in our definition of the dislocation elastic energy (${\bm u}_n=0$ $\forall n\geq1$)
and take only the Volterra solution.
The elastic field creates nevertheless tractions on the surface which isolates the dislocation core.
As a consequence, a contribution of the core tractions to the dislocation elastic energy exists
even when the dislocation core field is neglected. 
As it will be shown below, this contribution needs to be considered 
so as to obtain an unambiguous definition of the dislocation elastic energy.

\begin{figure}[hbtp]
	\begin{center}
		\includegraphics[width=0.9\linewidth]{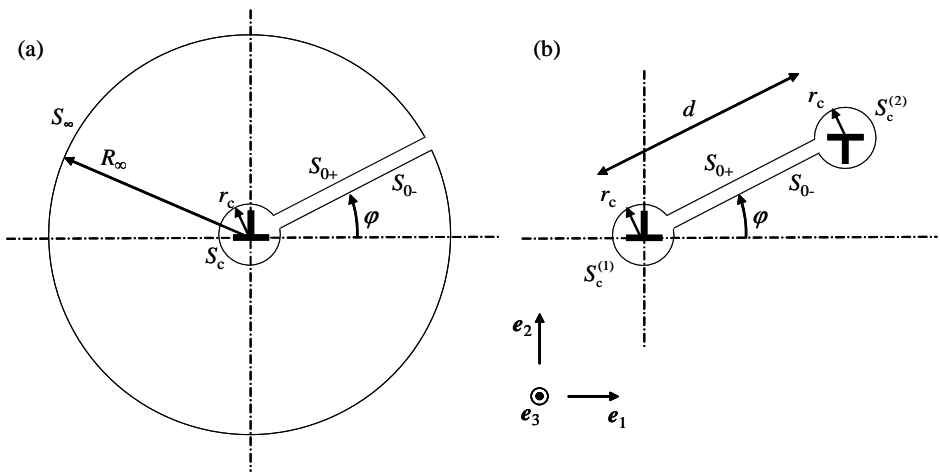}
	\end{center}
	\caption{Definition of the contour used to calculate the dislocation
		elastic energy.
		(a) Isolated dislocation.
		(b) Isolated dislocation dipole.}
	\label{fig:dislo}
\end{figure}

We first consider an isolated dislocation. 
Two cylinders centered on the dislocation need to be introduced so as to define 
the elastic energy (Fig.~\ref{fig:dislo}a).
The inner cylinder of surface $S_{\textrm{c}}$ isolates
the dislocation core. 
Strains are much too high close to the dislocation core to be described
by elasticity theory. As a consequence, elastic fields are diverging at the origin
and one needs to exclude the core region from the elastic description.
As the elastic energy integrated on an infinite volume tends to infinity, 
one also needs the external cylinder to prevent the elastic energy 
from diverging.
The dislocation elastic energy per unit-length of dislocation
is thus defined in the volume $V$ comprised 
between both cylinders
\begin{equation*}
	E^{\textrm{elas}}_{\rm dislo} = \frac{1}{2}\iiint_V{\sigma_{ij}\varepsilon_{ij}\md{V}},
\end{equation*}
where $\boldsymbol{\usigma}$ and $\boldsymbol{\uvarepsilon}$ are respectively
the stress and strain created by the dislocation.
The use of Gauss theorem allows to transform this volume integral into a surface integral:
\begin{equation}
	E^{\textrm{elas}}_{\rm dislo} = \frac{1}{2} \iint_S{\sigma_{ij}u_i\md{S_j}},
	\label{eq:elas_energy_S}
\end{equation}
where $S$ is a surface enclosing the volume $V$ and isolating any discontinuity
of the elastic displacement ${\bm u}$ generated by the dislocation.
A dislocation is by definition the frontier of a surface that has been sheared
by the Burgers vector ${\bm b}$. A displacement discontinuity therefore occurs on 
the dislocation cut which needs to be isolated by the surfaces $S_{0^+}$
and $S_{0^-}$ (Fig.~\ref{fig:dislo}a). 
We would like to stress that this cut does not necessary correspond to the
dislocation glide plane.
The surface appearing in the integral \ref{eq:elas_energy_S}
is then composed of $S_{\textrm{c}}$, $S_{0^-}$, $S_{\infty}$, and $S_{0^+}$.

The integrals along the surfaces $S_{\rm c}$ and $S_{\infty}$ cancel
because the resultant of forces located in the core is null, 
and the normals to the surfaces $S_{\rm c}$ and $S_{\infty}$
have an opposite orientation \cite{STR58,STR62}. 
As a consequence, the elastic energy of an isolated dislocation 
integrated between the cylinders of radii $r_{\rm c}$ and $R_{\infty}$
is limited to the contribution corresponding to the dislocation cut,
leading to the the well-known result
\begin{equation}
	E^{\rm elas}_{\rm dislo}
		=  \frac{1}{2} b_i K^0_{ij} b_j \ln{\left( \frac{R_{\infty}}{r_{\rm c}} \right)},
	\label{eq:Eelas_dislo}
\end{equation}
where the tensor ${\sf{\textbf K}}^0$ only depends on the elastic constants,
and $r_{\rm c}$ and $R_{\infty}$ are respectively the radii of the inner
and external cylinders.

We consider then an isolated dislocation dipole.
The first dislocation of Burgers vector ${\bm b}$ is located at the origin
and the second one of Burgers vector $-{\bm b}$ at the point
defined by its cylindrical coordinates $(d,\phi)$ (Fig.~\ref{fig:dislo}b).
The two dislocation cuts need to be orientated so that the displacement
discontinuity cancel except in the surface bounded by both dislocations.
This ensures that no displacement discontinuity occurs far from the dipole
and allows the dipole to be mechanically isolated.
The elastic energy created by the dipole in the infinite volume is then
\begin{equation*}
	E^{\rm elas}_{\rm dipole} = \frac{1}{2} \iint_{S}{
		\left( \sigma_{ij}^{(1)} + \sigma_{ij}^{(2)} \right)
		\left( u_i^{(1)} + u_i^{(2)} \right) \md{S_j} } ,
\end{equation*}
where ${\bm \sigma}^{(1)}$ and ${\bm \sigma}^{(2)}$ are the stresses 
created by each dislocation, and ${\bm u}^{(1)}$ and ${\bm u}^{(2)}$
the corresponding elastic displacements.
The integration surface is composed of the two cylinders $S_{\rm c}^{(1)}$ 
and $S_{\rm c}^{(2)}$ of radii $r_{\rm c}$ 
removing the elastic divergence at the dislocation cores, 
and of the two surfaces $S_{0^-}$ and $S_{0^+}$ removing the 
displacement discontinuity along the dislocation cut (Fig.~\ref{fig:dislo}b).
One does not need to introduce an external surface like for the isolated dislocation
because the elastic energy integrated on the infinite volume now converges.

The integration on both core cylinders leads to the same contribution
\begin{equation}
	E^{\rm elas}_{\rm c}(\phi) = \frac{1}{2} \iint_{S_{\rm c}^{(1)}} \sigma_{ij}^{(1)} u_i^{(1)} \md{S_j} 
	= \frac{1}{2} \iint_{S_{\rm c}^{(2)}} \sigma_{ij}^{(2)} u_i^{(2)} \md{S_j} .
	\label{eq:Ec1}
\end{equation}
The elastic energy of the dislocation dipole is then
\begin{equation}
	E^{\rm elas}_{\rm dipole} = 2 E^{\rm elas}_{\rm c}(\phi) 
		+  b_i K_{ij}^0 b_j \ln{\left( \frac{d}{r_{\rm c}} \right)} .
	\label{eq:Eelas_dipole}
\end{equation}

We will show in the following that the contribution $E^{\rm elas}_{\rm c}$
of the core tractions only depends on the angle $\phi$
defining the azimuthal position of the dislocation dipole.
It does not depend on the core radius $r_{\rm c}$
nor on the separation distance $d$.
This contribution is not present in the elastic energy of an isolated dislocation 
(Eq.~\ref{eq:Eelas_dislo}). This is a consequence of the introduction
of an external cylinder to prevent the elastic energy from diverging. 
But an isolated dislocation cannot exist: another defect, like a dislocation
with opposite Burgers vector or a surface, is always needed to close 
the dislocation cut and allows mechanical equilibrium. 
Therefore, one has to use the ``trick'' of the external cylinder 
so as to define the elastic energy of an isolated dislocation. 
This gives birth to an artefact as the core traction contribution
$E^{\rm elas}_{\rm c}$ then disappears.

\section{Isotropic elastic media}

We first consider that the elastic constants are isotropic. 
This allows to obtain simple expressions of the elastic energy
which can be easily manipulated so as to illustrate
the importance of the core traction contribution.
We assume that the crystal is oriented in such a way
that ${\bm e}_3$ corresponds to the dislocation axis
and ${\bm e}_1$ is collinear with the edge component of the Burgers
vector ($b_2=0$). 
With such an orientation, the tensor ${\sf{\textbf K}}^0$ appearing
in the elastic energy of an isolated dislocation (Eq.~\ref{eq:Eelas_dislo})
or of a dislocation dipole (Eq.~\ref{eq:Eelas_dipole}) writes \cite{HIR82}
\begin{equation*}
	{\sf{\textbf K}}^0 = \frac{\mu}{2\upi\left( 1-\nu \right)} 
	\begin{pmatrix}
		\phantom{(1}1\phantom{\nu)} & 0 & 0 \\
		0 & \phantom{(1}1\phantom{\nu)} & 0 \\
		0 & 0 & \left( 1-\nu \right) 
	\end{pmatrix} ,
\end{equation*}
where $\mu$ is the shear modulus and $\nu$ Poisson's ratio.

The core traction contribution has been calculated
by Bullough and Foreman \cite{BUL64}:
\begin{equation*}
	E^{\rm elas}_{\rm c}(\phi) = \frac{\mu {b_1}^2}{16\upi\left( 1-\nu \right)}
	\left[ \frac{1}{1-\nu} - 2\cos{\left( 2\phi \right)} \right].
\end{equation*}
One sees that this contribution is null for a pure screw dislocation ($b_1=0$), 
but this result does not hold anymore when elastic anisotropy is considered 
(§~\ref{sec:elasticity_ani}).
For a pure edge or a mixed dislocation, if the dislocation cut corresponds
to the dislocation glide plane, one has then $\phi=0$ or $\phi=\upi$,
and one recovers the expression given by Hirth and Lothe (Eq.~(3-53) in Ref.~\cite{HIR82}):
\begin{equation}
	E^{\rm elas}_{\rm c}\left( \phi=0 \right) = - \frac{\mu {b_1}^2}{16\upi}
	\frac{1-2\nu}{\left(1-\nu\right)^2} .
	\label{eq:Ec_iso_edge}
\end{equation}
As Poisson's ratio is smaller than $1/2$, this shows that 
the core traction contribution reduces the dislocation elastic energy
when the dislocation cut corresponds to its glide plane.

\begin{figure}[hbtp]
	\begin{center}
		\includegraphics[width=0.3\linewidth]{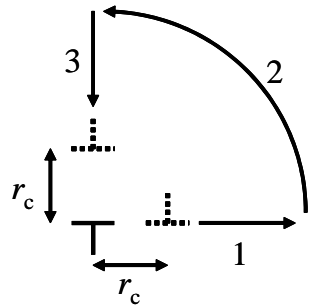}
	\end{center}
	\caption{Reversible thermodynamic cycle for a dipole 
	of edge dislocations.}
	\label{fig:dipole_cycle}
\end{figure}

To understand the importance of the core traction contribution, 
one can look at the variation of energy for a dipole of edge 
dislocations subjected to the cycle sketched in Fig.~\ref{fig:dipole_cycle}.
This cycle has been proposed by Hirth and Lothe (Ref.~\cite{LOT05}
and section 8-3 in Ref.~\cite{HIR82}).
The dipole is first created with both edge dislocations lying 
on the same glide plane and separated by a distance $r_{\rm c}$.
The energy required to create the dipole is $2E^{\rm core} 
+ 2E^{\rm elas}_{\rm c}\left( \phi=0 \right)$, 
where $E^{\rm core}$ is the dislocation core energy,
\ie the part of energy which cannot be described by linear elasticity.
One dislocation is then displaced through the cycle $1\to2\to3$.
The work performed by the Peach-Koehler force during this displacement
is $\mu b^2\left( 1-\nu \right)/{2\upi}$.
When the dislocations are at a distance $r_{\rm c}$, 
the dipole is destroyed and one recovers the energy
$2E^{\rm core} + 2E^{\rm elas}_{\rm c}\left( \phi=\upi/2 \right)$.
The variation of energy through the complete cycle is then
$2E^{\rm elas}_{\rm c}\left( \phi=\upi/2 \right) 
- 2E^{\rm elas}_{\rm c}\left( \phi=0 \right)
+\mu b^2\left( 1-\nu \right)/{2\upi} = 0$.
If one had neglected the core traction contribution
to the dipole elastic energy, the energy variation 
would have been non-null which violates the law of thermodynamics.
The proper consideration of all energy contributions 
in the Volterra elastic field thus allows a coherent 
definition of dislocation energetics. 
Such a coherency does not require more complex descriptions
which assume a spreading of the dislocation like in the Peierls-Nabarro 
model or in the standard core model proposed by Lothe \cite{LOT92}.

\section{Anisotropic elastic media}
\label{sec:elasticity_ani}

\subsection{Sextic formalism}

We consider a dislocation of Burgers vector ${\bm b}$ located at the origin
and we orient the crystal so that the axis ${\bm e}_3$ corresponds to the
dislocation line.
We assume that the angle $\phi$ in Fig.~\ref{fig:dislo} is equal to $-\upi$:
the dislocation cut corresponds with the half plane defined by $x_1<0$
and $x_2=0$.
We will generalize in a second stage to a different angle $\phi$.

The elastic displacement and the elastic stress created by the dislocation
are given to the first order by the Volterra solution. 
They can be expressed using the formalism\footnote{For a more detailled presentation
of the sextic formalism, \cf chap. 13 in Ref. \cite{HIR82} of Refs. \cite{BAC80} and \cite{TIN96}.}
developed by Eshelby \etal \cite{ESH53}
and extended by Stroh \cite{STR58,STR62}.
The displacement and the stress calculated in a point of Cartesian coordinates
$\left( x_1, x_2, x_3 \right)$ are then
\begin{equation}
\begin{split}
	u_k(x_1,x_2) &= \frac{1}{2} \sum_{\alpha=1}^{6}{
		\mp\frac{1}{2\upi\mi} A_k^{\alpha} 
		D_{\alpha} \log{\left( x_1 + p_{\alpha} x_2 \right)} } \\
	\sigma_{ij}(x_1,x_2) &= \frac{1}{2} \sum_{\alpha=1}^{6}{
		\mp\frac{1}{2\upi\mi} B_{ijk}^{\alpha} A_k^{\alpha} 
		D_{\alpha} \frac{1}{ x_1 + p_{\alpha} x_2 }} 
\end{split}
\label{eq:elast_esh}
\end{equation}
The sign $\mp$ appearing in these equations\footnote{We use the same sign
convention as Hirth and Lothe \cite{HIR82}. Eshelby \etal \cite{ESH53}
and Stroh \cite{STR58,STR62} use the opposite sign in Eq.~\ref{eq:elast_esh}.}
means $-$ for $1\leq\alpha\leq3$ and $+$ for $4\leq\alpha\leq6$.

The matrices $B_{ijk}^{\alpha}$ are obtained from the elastic constants $C_{ijkl}$
expressed in the dislocation reference frame:
\begin{equation*}
	B_{ijk}^{\alpha} = C_{ijk1} + p_{\alpha} C_{ijk2}.
\end{equation*}
The roots $p_{\alpha}$ are solution of the sextic equations 
corresponding to the following determinant being equal to zero
\begin{equation*}
	\left|\left\{ B_{i1k}^{\alpha}+p_{\alpha}B_{i2k}^{\alpha} \right\}\right| = 0,	
\end{equation*}
and the non-null vectors $A_k^{\alpha}$, associated to each root $p_{\alpha}$,
check the following equation
\begin{equation}
	(B_{i1k}^{\alpha}+p_{\alpha}B_{i2k}^{\alpha}) A_k^{\alpha} = 0.
	\label{eq:sextic}
\end{equation}
In all equations, we use the Einstein summation convention on repeated indexes,
except for indexes written in Greek letters.
When summation on the six different roots $p_{\alpha}$ is required, 
it will be explicitly written like in Eq.~\ref{eq:elast_esh}.

The six roots $p_{\alpha}$ are necessary complex.
If $p_{\alpha}$ is solution of Eq.~\ref{eq:sextic}, 
its complex conjugate ${p_{\alpha}}^*$ is also a solution
of this equation. We can therefore assume that the six different
roots have been sorted in such a way that
\begin{equation}
	\Im{(p_{\alpha})} > 0 \textrm{ and } p_{\alpha+3}={p_{\alpha}}^*, \quad 1\leq\alpha\leq3,
	\label{eq:p_cc}
\end{equation}
where $\Im{(p_{\alpha})}$ is the imaginary part of $p_{\alpha}$.

The elastic field given by Eq.~\ref{eq:elast_esh} corresponds to the one 
of a dislocation with a Burgers vector ${\bm b}$ and a line direction ${\bm e}_3$
if the constants $D_{\alpha}$ check the equations:
\begin{equation}
	\begin{split}
		\frac{1}{2} \sum_{\alpha=1}^{6}{A_k^{\alpha} D_{\alpha}} &= -b_k , \\
		\frac{1}{2} \sum_{\alpha=1}^{6}{ B_{i2k}^{\alpha} A_k^{\alpha} D_{\alpha}} &= 0 .
	\end{split}
	\label{eq:D_eshelby}
\end{equation}
We choose the principal determination for the complex logarithm 
appearing in Eq.~\ref{eq:elast_esh}. The elastic 
displacement created by the dislocation therefore presents a 
discontinuity in the half plane of equations $x_1<0$ and $x_2=0$ 
defining the dislocation cut.

The system of linear equations \ref{eq:D_eshelby} can be easily solved
following Stroh method \cite{STR58,STR62}. 
To do so, we define a new vector
\begin{equation}
	L_i^{\alpha} = B_{i2k}^{\alpha}A_k^{\alpha} 
	             = -\frac{1}{p_{\alpha}}B_{i1k}^{\alpha}A_k^{\alpha} .
	\label{eq:L_Stroh}
\end{equation}
Both definitions are equivalent because of Eq.~\ref{eq:sextic}.
As the vector $A_k^{\alpha}$ is an eigenvector defined by Eq.~\ref{eq:sextic},
its norm is not fixed. One can therefore choose it so as to check 
the following normalization condition:
\begin{equation*}
	2 A_i^{\alpha} L_i^{\alpha} = 1, \quad \forall \alpha.
\end{equation*}
Using the orthogonality properties \cite{STR58,STR62,BAC80,TIN96} of the vectors $A_i^{\alpha}$ and $L_i^{\alpha}$,
the solution of the system of equations \ref{eq:D_eshelby} is given by
\begin{equation}
	D_{\alpha} = -2 L_i^{\alpha} b_i .
	\label{eq:D_Stroh}
\end{equation}

With such definitions, the tensor $\sf{\textbf{K}}^{0}$ appearing in the elastic energy
of an isolated dislocation (Eq.~\ref{eq:Eelas_dislo}) 
or of a dislocation dipole (Eq.~\ref{eq:Eelas_dipole}) is given by
\begin{equation}
	K^0_{ij}  = \sum_{\alpha=1}^6{ \pm\frac{1}{2\upi\mi} L_i^{\alpha}L_j^{\alpha} }	.
	\label{eq:Kij_dislo}
\end{equation}

\subsection{The core tractions contribution}

The contribution of the core tractions to the elastic energy (Eq.~\ref{eq:Ec1})
is given by
\begin{equation}
	E^{\rm elas}_{\rm c} = -\frac{1}{2} \int_{\theta=-\upi}^{\upi}{
		\left[ \sigma_{i1}(r_{\rm c},\theta)\cos{(\theta)}
			+ \sigma_{i2}(r_{\rm c},\theta)\sin{(\theta)}\right]
		u_i(r_{\rm c},\theta) r_{\rm c} \md{\theta}}.
	\label{eq:Ec2}
\end{equation}
The expression \ref{eq:elast_esh} shows that the elastic displacement
${\bm u}(r_{\rm},\theta)$ is the sum of an angular function
and a term depending only on $\ln{\left( r_{\rm c} \right)}$. 
This last term leads to a contribution in Eq.~\ref{eq:Ec2} which is null
because the resultant of forces located in the core 
is null for a pure dislocation
\cite{STR58,STR62}. 
Only the angular part of the elastic displacement leads to a
contribution in Eq. \ref{eq:Ec2}. One thus obtains
\begin{multline*}
	E_{\rm c}^{\rm elas} = -\frac{1}{8}\sum_{\alpha=1}^6{\mp\frac{1}{2\upi\mi}A_i^{\alpha}D_{\alpha}
	\sum_{\beta=1}^6{\mp\frac{1}{2\upi\mi}A_k^{\beta}D_{\beta} }} \\
	\int_{-\upi}^{\upi}{\frac{B_{i1k}^{\beta}\cos{(\theta)}+B_{i2k}^{\beta}\sin{(\theta)}}
	{\cos{(\theta)}+p_{\beta}\sin{(\theta)}}
	\log{\left[\cos{(\theta)}+p_{\alpha}\sin{(\theta)}\right]}\md{\theta}}.
\end{multline*}
This expression already shows that the contribution of the core tractions
to the elastic energy does not depend on the radius $r_{\rm c}$ of the 
core cylinder.
Using the property \ref{eq:sextic} of the matrices $B_{ijk}^{\beta}$
and the vectors $A_k^{\beta}$, as well as the definition \ref{eq:L_Stroh}
of the vector $L_i^{\beta}$, one gets
\begin{equation}
	E_{\rm c}^{\rm elas} = \frac{1}{8}\sum_{\alpha=1}^6{\pm\frac{1}{2\upi\mi} 
	\sum_{\beta=1}^6{\pm\frac{1}{2\upi\mi}
	D_{\alpha} A_i^{\alpha} L_i^{\beta} D_{\beta}
	J^1(p_{\alpha},p_{\beta})}} ,
	\label{eq:Ec3}
\end{equation}
where the integral $J^1(p,a)$ is defined by
\begin{equation}
	J^1(p,q) = \int_{-\upi}^{\upi}{\frac{-p\cos{\theta}+\sin{\theta}}
	{\cos{\theta}+p\sin{\theta}} 
	\log{\left( \cos{\theta}+q\sin{\theta} \right)}\md{\theta}}.
	\label{eq:J1}
\end{equation}
An analytical expression of this integral is obtained in the appendix
\ref{sec:integral_log_J1}. Using this expression 
with the properties \ref{eq:p_cc}
checked by the roots $p_{\alpha}$, one obtains
\begin{multline}
	E_{\rm c}^{\rm elas} = \frac{1}{8}\sum_{\alpha=1}^6{
	\log{(\mi \pm p_{\alpha})} 
	\sum_{\beta=1}^6{
	\pm\frac{1}{2\upi\mi}D_{\alpha}
	\left( A_i^{\alpha}L_i^{\beta}-L_i^{\alpha}A_i^{\beta} \right)
	D_{\beta}}} \\
	+ \frac{1}{8\upi\mi}\sum_{\alpha=1}^3{\sum_{\beta=4}^6{
	D_{\alpha}
	\left( A_i^{\alpha}L_i^{\beta}-L_i^{\alpha}A_i^{\beta} \right)
	D_{\beta} \log{\left( p_{\alpha}-p_{\beta} \right)} }} .
	\label{eq:energy_dislo_core}
\end{multline}
We therefore obtained an analytical expression of the core traction contribution
within the sextic formalism.

\subsection{Angular dependence of the dipole elastic energy}

We now examine how the dipole elastic energy varies when 
it is rotated with respect to the crystallographic axes.
The answer could be obtained by rotating the dislocations 
and the elastic constants so as to calculate in the new 
reference frame all quantities needed to express the 
elastic energy. But a closed-form
expression of the angular dependence of the elastic energy
can be obtained in a fixed reference frame. 
We will show that this expression is consistent 
with the dislocation interaction energy derived by Stroh \cite{STR58}.

The angle $\phi$ can now deviate from the value $-\upi$ considered above.
A rotation of the dislocation dipole is equivalent to a rotation
of the corresponding cut (Fig.~\ref{fig:dislo}). 
Such a rotation does not modify the vectors $A_i^{\alpha}$
and $L_i^{\alpha}$ \cite{BAC80,TIN96}. As a consequence, the tensor
${\sf{\textbf K}}^0$ appearing in the dipole elastic energy (Eq.~\ref{eq:Eelas_dipole})
is unchanged.
Only the contribution $E^{\rm elas}_{\rm c}$ of the dislocation core tractions
depends on this angle $\phi$. 
We then call $\Delta E^{\rm elas}_{\rm c}(\phi)$ the variation of the dipole
elastic energy, with the convention $\Delta E^{\rm elas}_{\rm c}(-\upi)=0$.
To calculate this energy variation, it is useful to write 
the roots of the sextic equations in the form 
$p_{\alpha}=\tan{(\psi_{\alpha})}$. 
Such a transformation can be performed as long as the roots
differ from $\pm\mi$ which only happens in degenerate cases
due to some isotropy of the elastic constants\footnote{The degeneracy
can be lifted by adding some noise to the elastic constants.} \cite{BAC80}.
We can now rewrite the integral (Eq.~\ref{eq:J1}) appearing in 
the energy contribution of the core tractions (Eq.~\ref{eq:Ec3})
\begin{equation*}
	J^1(p_{\alpha},p_{\beta}) = \int_{-\upi}^{\upi}{
	-\tan{(\psi_{\alpha}-\theta)}
	\log{\left[ \frac{ \cos{(\psi_{\beta}-\theta)} }{\cos{(\psi_{\beta})}} \right]}
	\md{\theta}}.
\end{equation*}
The rotation of the cut modify the roots through the relation
$p_{\alpha}(\phi)=\tan{(\psi_{\alpha}-\phi)}$ \cite{BAC80}.
As a consequence, the elastic energy variation is given by
\begin{equation*}
	\Delta{E_{\rm c}^{\rm elas}}(\phi) = \frac{1}{8}\sum_{\alpha=1}^6{\pm\frac{1}{2\upi\mi} 
	\sum_{\beta=1}^6{\pm\frac{1}{2\upi\mi}
	D_{\alpha} A_i^{\alpha} L_i^{\beta} D_{\beta}
	\Delta{J^1_{\phi}}(p_{\alpha},p_{\beta})}} ,
\end{equation*}
with
\begin{align*}
	\Delta{J^1_{\phi}}(p_{\alpha},p_{\beta}) 
	=& \int_{-\upi}^{\upi}
	-\tan{(\psi_{\alpha}-\theta)}  \left\{ \log{\left[ \frac{ \cos{(\psi_{\beta}-\theta)} }
			{\cos{(\psi_{\beta}-\phi)}} \right]}
	- \log{\left[ \frac{ \cos{(\psi_{\beta}-\theta)} }
			{\cos{(\psi_{\beta})}} \right]} \right\}
	\md{\theta} \\
	=& \int_{-\upi}^{\upi}
	\frac{-p_{\alpha}\cos{\theta}+\sin{\theta}}
	{\cos{\theta}+p_{\alpha}\sin{\theta}} \\
	& \qquad \left[ \log{\left( \frac{ \cos{\theta}+p_{\beta}\sin{\theta} }
	{\cos{\phi}+p_{\beta}\sin{\phi}}\right)}
	 - \log{\left( \cos{\theta}+p_{\beta}\sin{\theta} \right)}
	\right]\md{\theta}.
\end{align*}
This integral is calculated in the appendix \ref{sec:integral_log_dJ1}.
With the result of this appendix, one obtains
\begin{align*}
	\Delta{E_{\rm c}^{\rm elas}}(\phi) =&
	\frac{1}{8}\sum_{\alpha=1}^6{\sum_{\beta=1}^6{\pm\frac{1}{2\upi\mi}
	D_{\alpha} A_i^{\alpha} L_i^{\beta} D_{\beta}
	\log{(\cos{\phi}+p_{\beta}\sin{\phi})} }} \\
	&- \frac{1}{8}\sum_{\alpha=1}^6{\pm\frac{1}{2\upi\mi}\sum_{\beta=1}^6{
	D_{\alpha} A_i^{\alpha} L_i^{\beta} D_{\beta}
	\log{(\cos{\phi}+p_{\alpha}\sin{\phi})} }} .
\end{align*}
Using the orthogonality properties checked by the vectors
$A_i^{\alpha}$ and $L_i^{\alpha}$, 
as well as the definition \ref{eq:D_Stroh} of $D_{\alpha}$,
one gets
\begin{equation}
	\Delta{E_{\rm c}^{\rm elas}}(\phi) =
	\frac{1}{2}\sum_{\alpha=1}^6{\pm\frac{1}{2\upi\mi}
	b_i L_i^{\alpha} L_j^{\alpha} b_j 
	\log{(\cos{\phi}+p_{\alpha}\sin{\phi})}}.
	\label{eq:Ec_angular}
\end{equation}

One recovers an angular dependence which is in agreement
with the expression of the interaction energy between two 
dislocations given by Stroh \cite{STR58}, as well as by
Cai \etal \cite{CAI03}.
\footnote{The expression used by Li \etal \cite{LI04},
based on the work of Cai \etal \cite{CAI03},
is different from Eq.~\ref{eq:Ec_angular}.
Li obtained $\Delta{E^c}(\phi) = \frac{1}{2\upi} \sum_{\alpha=1}^3{
\Im{ \left\{ b_iL_i^{\alpha}L_j^{\alpha}b_j  \right\}}
\Re{\left\{ \log{(\cos{\phi}+p_{\alpha}\sin{\phi})} \right\} } }$,
whereas Eq.~\ref{eq:Ec_angular} can be rewritten
$\Delta{E^c}(\phi) = \frac{1}{2\upi} \sum_{\alpha=1}^3{
\Im{ \left\{ b_iL_i^{\alpha}L_j^{\alpha}b_j  
\log{(\cos{\phi}+p_{\alpha}\sin{\phi})} \right\} } }$.
Both definitions may be equivalent in the case of a screw dislocation
studied by Li \etal, but we could not demonstrate it.}

This expression \ref{eq:Ec_angular} shows too that the core
traction contribution $E_{\rm c}^{\rm elas}$ is periodic 
of period $\upi$, \ie is insensitive to an inversion 
of the cut direction. 
We have indeed
\begin{equation*}
	\Delta{E_{\rm c}^{\rm elas}}(\phi + \upi) = 
	\frac{1}{2}\sum_{\alpha=1}^6{\pm\frac{1}{2\upi\mi}
	b_i L_i^{\alpha} L_j^{\alpha} b_j 
	\log{(-\cos{\phi}-p_{\alpha}\sin{\phi})}} .
\end{equation*}
Using the property of $\log{\left( -z \right)}$, we obtain
\begin{align*}
	\Delta{E_{\rm c}^{\rm elas}}(\phi + \upi) 
	= & \Delta{E_{\rm c}^{\rm elas}}(\phi)
	- \frac{1}{4}\sum_{\alpha=1}^6{ b_i L_i^{\alpha} L_j^{\alpha} b_j }
	\quad \textrm{ if }0<\phi<\upi \\
	= & \Delta{E_{\rm c}^{\rm elas}}(\phi)
	+ \frac{1}{4}\sum_{\alpha=1}^6{ b_i L_i^{\alpha} L_j^{\alpha} b_j }
	\quad \textrm{ if }-\upi<\phi<0.
\end{align*}
Finally, the closure properties of the vectors $L_i^{\alpha}$
\cite{STR62,BAC80,TIN96} lead to the desired result:
\begin{equation*}
	\Delta{E_{\rm c}^{\rm elas}}(\phi + \upi) 
	=  \Delta{E_{\rm c}^{\rm elas}}(\phi).
\end{equation*}

\section{Edge dislocations in $\alpha$-iron}

So as to illustrate the need to take into account the contribution
of core tractions to the dislocation elastic energy, we study now
edge dislocations in $\alpha$-iron, and determine their core
energies $E^{\rm core}$. The core energy is the amount of the dislocation excess energy 
that arises from atomic interaction close to the dislocation core
that cannot be described by linear elasticity.
This is therefore a supplementary energy contribution that should be 
added to the elastic energy. One expects that such an energy contribution
is an intrinsic property of the dislocation: its value should only depend
on the dislocation and not on the surrounding environment, like other
dislocations. 

We use the Fe empirical potential developed by Mendelev \etal \cite{MEN03}
in its modified version published in Ref.~\cite{ACK04}.
Thanks to the existence of a cut-off
for the interactions between atoms, two different methods can be used 
to determine the dislocation core energy. One can either work 
with an isolated dislocation in an infinite elastic medium (the cluster
approach), or with a dislocation dipole in periodic boundary
conditions (the dipole approach). In the following section, 
we will show that both methods are coherent and lead to the same
core energy as long as all contributions are considered in the
elastic energy.

$\alpha$-Fe has a body-centered cubic lattice and the Burgers vector
of the most common dislocations is ${\bm b}=a/2\left<111\right>$,
where $a=2.8553$~{\AA} is the lattice vector. 
Depending on the temperature, different slip planes are observed,
either $\left\{ 110 \right\}$ or $\left\{ 112 \right\}$ planes.
It is therefore interesting to study the energetics 
of both $1/2\left<111\right>\left\{110\right\}$ and 
$1/2\left<111\right>\left\{112\right\}$ edge dislocations, 
which both can exist. We will therefore determine the core
energy of these two dislocations.
For the $1/2\left<111\right>\left\{110\right\}$ edge dislocation,
we use a simulation box with the orientation
${\bm e}_1=\left[ 111 \right]/\sqrt{3}$, 
${\bm e}_2=\left[ \bar{1}01 \right]/\sqrt{2}$, 
and ${\bm e}_3=\left[ 1\bar{2}1 \right]/\sqrt{6}$; 
for the $1/2\left<111\right>\left\{112\right\}$ edge dislocation
${\bm e}_1=\left[ 111 \right]/\sqrt{3}$, 
${\bm e}_2=\left[ \bar{1}2\bar{1} \right]/\sqrt{6}$, 
and ${\bm e}_3=\left[ \bar{1}01 \right]/\sqrt{2}$. 
In both cases, ${\bm e}_1$ is collinear with the Burgers vector
and ${\bm e}_3$ corresponds to the dislocation line. 
The dislocation glide plane is therefore the plane of normal 
${\bm e}_2$ ($\phi=0$ in Fig.~\ref{fig:dislo}).

\subsection{The cluster approach}

In the cluster approach, a single dislocation is introduced
at the center of a unit cell which is periodic only along
the dislocation line and with surface in other directions. 
Atoms are displaced according to the anisotropic elastic 
Volterra displacement (Eq.~\ref{eq:elast_esh}). 
Atoms closer from the external surface than the interatomic
potential cut-off are kept fixed while the positions of inner atoms 
are relaxed so as to minimize the energy calculated 
with the potential.
This therefore simulates an isolated dislocation 
in an infinite elastic medium.
A variant of the method consists in relaxing atoms at the surface
using lattice Green functions \cite{SIN78,WOO01b,WOO02}.
This may be necessary in ab-initio calculations because of the
small size of the unit cell that can be simulated. 
As, in the present work, we use empirical potential,
the unit cell is large enough so that the Volterra elastic field 
correctly describes the displacements of atoms at the surface 
and one does not need to use lattice Green function to relax them.

\begin{figure}[bhtp]
	\begin{center}
		\subfigure[$1/2\left<111\right>\left\{110\right\}$]{
			\includegraphics[width=0.47\linewidth]{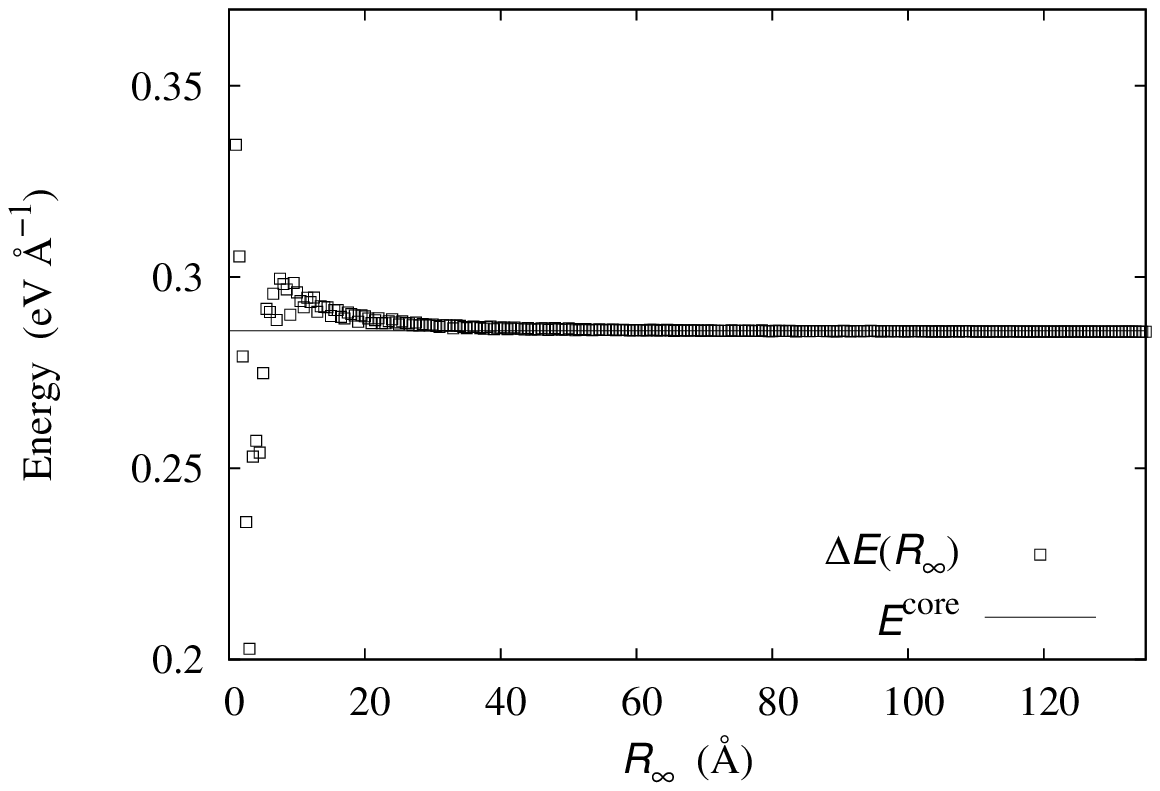}}
		\hfill
		\subfigure[$1/2\left<111\right>\left\{112\right\}$]{
			\includegraphics[width=0.47\linewidth]{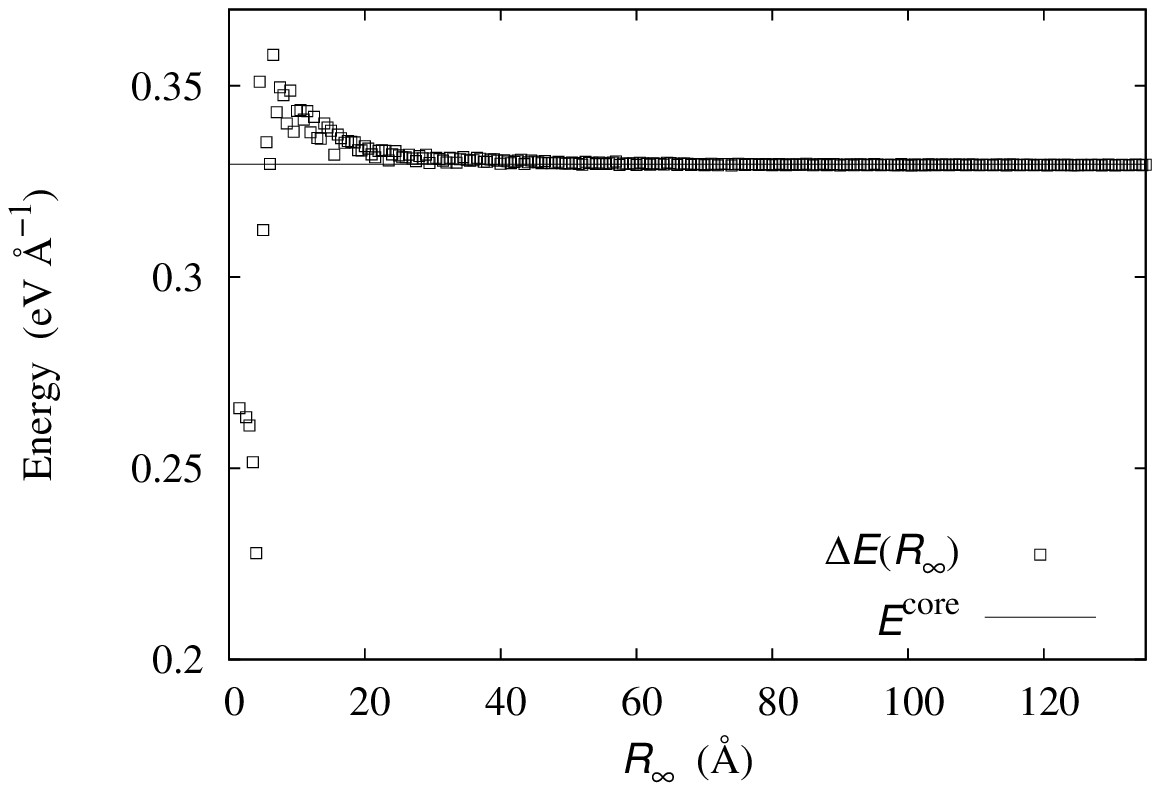}}
	\end{center}
	\caption{Variation with the cylinder radius $R_{\infty}$ of the 
	energy difference $\Delta E\left( R_{\infty} \right) =
	E\left( R_{\infty} \right) - \frac{1}{2} b_iK^0_{ij}b_j
	\ln{\left(R_{\infty}/r_{\rm c}\right)}$ showing the convergence
	to $E^{\rm core}$ for $1/2\left<111\right>\left\{110\right\}$
	and $1/2\left<111\right>\left\{112\right\}$ edge dislocations
	in $\alpha$-iron ($r_{\rm c}=b\simeq2.473$~\AA).}
	\label{fig:Ecore_cluster}
\end{figure}

\begin{table}[bhtp]
	\tbl{Parameters defining the energies of 
	both edge dislocations
	in $\alpha$-iron. All quantities are given in meV~\AA$^{-1}$.
	The core energies are given for a core radius $r_{\rm c}=b$.}
	{\begin{tabular}{lccc}
		\toprule
		& $E^{\rm core}$	
		& $E^{\rm elas}_{\rm c}\left( \phi=0 \right)$
		& $\frac{1}{2}b_iK^0_{ij}b_j$ \\
		\colrule
		$1/2\left<111\right>\left\{110\right\}$ & 
			286 & $-$87 & 370 \\
		$1/2\left<111\right>\left\{112\right\}$ & 
			329.5 & $-$89 & 378 \\
		\botrule
	\end{tabular}}
	\label{tab:Ecore_Fe}
\end{table}

Once inner atoms are relaxed, we calculate the excess energy
$E\left( R_{\infty} \right)$ given by the interatomic potential and
contained in a cylinder centered on the dislocation, of axis ${\bm e}_3$
and of radius $R_{\infty}$. This excess energy is the difference 
of energy between the system with the dislocation and the perfect
crystal for the same number of atoms. Only the excess energy 
of atoms contained in the cylinder is considered.
This can be performed for different values of the radius $R_{\infty}$
in the limit of the maximal radius allowed by the unit cell. 
According to the section \ref{sec:energy} (Eq.~\ref{eq:Eelas_dislo}),
this excess energy should vary as
\begin{equation*}
	E\left( R_{\infty} \right) = E^{\rm core} 
	+ \frac{1}{2}b_iK^0_{ij}b_j\ln{\left( \frac{R_{\infty}}{r_{\rm c}} \right)}.
\end{equation*}
In this expression, the tensor ${\sf{\textbf K}}^0$ is given by 
Eq.~\ref{eq:Kij_dislo}. It is different for the 
$1/2\left<111\right>\left\{110\right\}$ and 
$1/2\left<111\right>\left\{112\right\}$ edge dislocations 
because of elastic anisotropy (Tab.~\ref{tab:Ecore_Fe}).
Looking at the variations of the energy difference
$\Delta E\left( R_{\infty} \right) = E\left( R_{\infty} \right) 
- \frac{1}{2} b_iK^0_{ij}b_j \ln{\left(R_{\infty}/r_{\rm c}\right)}$,
one expects this quantity to be a constant defining $E^{\rm core}$. 
This quantity actually varies but rapidly converges to a constant
value for an increasing radius $R_{\infty}$ (Fig.~\ref{fig:Ecore_cluster}). 
Variations for small $R_{\infty}$ arises from the dislocation core field,
which may be important close to the dislocation core 
but is not taken into account in the present approach.
This core field needs only to be considered when one cannot reach large enough simulation boxes
like in ab initio calculations \cite{CLO09}.
The convergence of $\Delta E\left( R_{\infty} \right)$ allows us 
anyway to define a dislocation core energy. Values of this core
energy are given for both dislocation in Tab.~\ref{tab:Ecore_Fe}.

\subsection{The dipole approach}

\begin{figure}[bhtp]
	\begin{center}
		\includegraphics[width=0.5\linewidth]{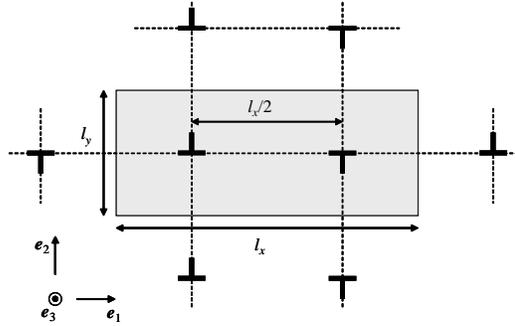}
	\end{center}
	\caption{Sketch of the periodic unit cell
	used to study a dipole of edge dislocations.}
	\label{fig:dipole_sketch}
\end{figure}

In the dipole approach, full periodic boundary conditions are used.
The total Burgers vector of the unit cell has to be zero.
Therefore a dislocation dipole is introduced in the unit cell.
The periodic unit cell we used for the present work is sketched in
Fig.~\ref{fig:dipole_sketch}. 
Both dislocations composing the dipole share the same glide plane
and then $\phi=0$.

\begin{figure}[bhtp]
	\begin{center}
		\subfigure[$1/2\left<111\right>\left\{110\right\}$]{
			\includegraphics[width=0.47\linewidth]{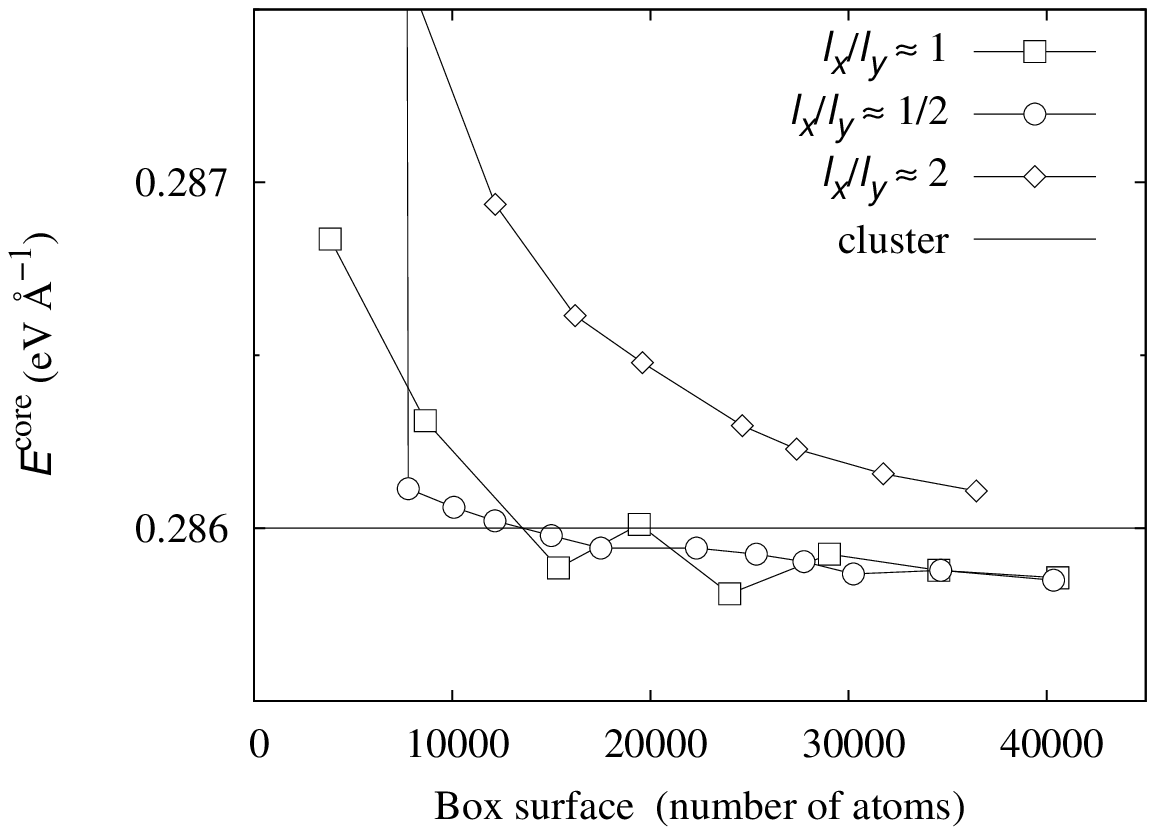}}
		\hfill
		\subfigure[$1/2\left<111\right>\left\{112\right\}$]{
			\includegraphics[width=0.47\linewidth]{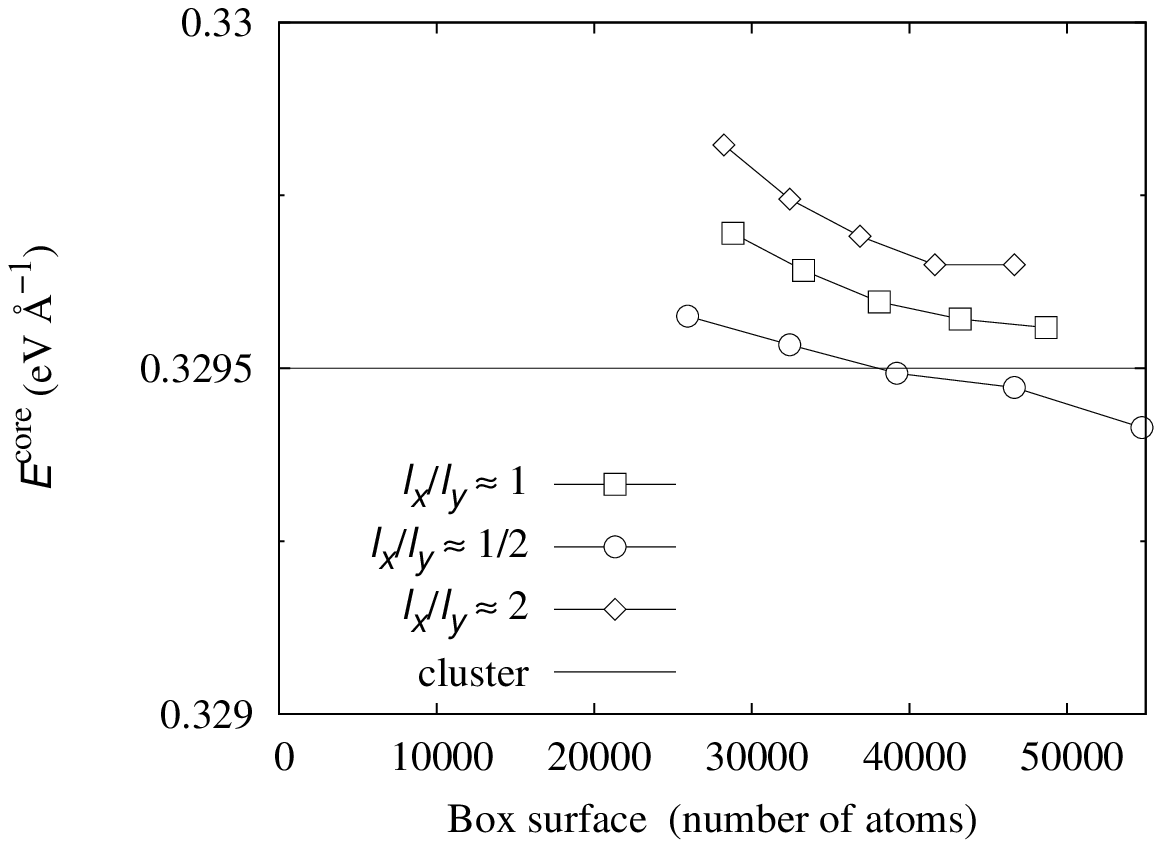}}
	\end{center}
	\caption{Variation with the size of the simulation cell 
	of the dislocation core energy deduced from atomic simulations
	of a dislocation dipole in periodic boundary conditions
	for $1/2\left<111\right>\left\{110\right\}$
	and $1/2\left<111\right>\left\{112\right\}$ edge dislocations
	in $\alpha$-iron ($r_{\rm c}=b\simeq2.473$~\AA).
	Symbols correspond to result of the dipole approach for different
	geometries of the unit cell
	and the horizontal line to the result of the cluster approach.}
	\label{fig:Ecore_dipole}
\end{figure}

The energy of the unit cell is minimized by relaxing all atomic 
positions. According to section \ref{sec:energy}, the excess
energy of the unit cell should be
\begin{equation}
	E = 2 E^{\rm core} + 2 E^{\rm elas}_{\rm c}\left( \phi=0 \right)
	+ b_i K^0_{ij} b_j \ln{\left( \frac{d}{r_{\rm c}} \right)}
	+ E^{\rm elas}_{\rm inter},
	\label{eq:Ecore_dipole}
\end{equation}
with $d=l_x/2$.
$E^{\rm elas}_{\rm inter}$ is the elastic interaction energy
between the dislocation dipole contained in the unit cell and all 
its periodic images. 
This elastic interaction energy is calculated using the method 
of Cai \etal \cite{CAI03}.
One can therefore deduce the core energy from the atomic simulations
using Eq.~\ref{eq:Ecore_dipole}: the excess energy $E$ is directly 
given by atomic simulations and all elastic contributions are calculated
using a code based on anisotropic linear elasticity.

The obtained core energies are presented in figure~\ref{fig:Ecore_dipole}
for different sizes of the unit cells as well as different geometries
characterized by the aspect ratio $l_x/l_y$.
All geometries converge with the size of the unit cell to the same limit
and the converged value of the core energy is in perfect agreement with 
the value obtained in the cluster approach.
The table \ref{tab:Ecore_Fe} shows that this agreement is possible
only if one does not forget the contribution of the core tractions 
to the elastic energy. Without this contribution, the core energies 
would have been about 90~meV~\AA$^{-1}$ lower in the dipole approach
than in the cluster approach.

\subsection{Stability of an edge dipole}

\begin{figure}[bhtp]
	\begin{center}
		\includegraphics[width=0.6\linewidth]{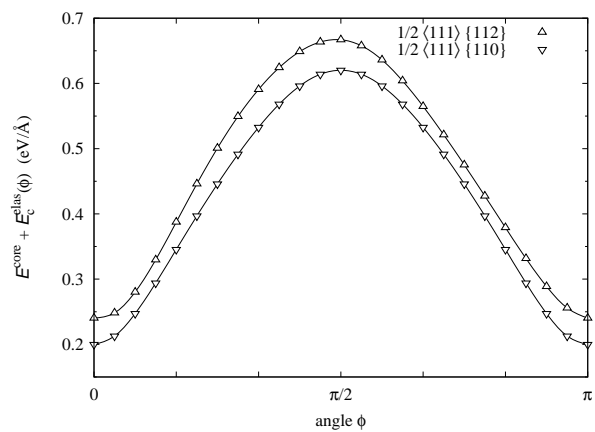}
	\end{center}
	\caption{Variation with the angle $\phi$ of the energy contribution
	$E^{\rm core}+E_{\rm c}^{\rm elas}(\phi)$ ($r_{\rm c}=b$).}
	\label{fig:Ephi}
\end{figure}

Eq. \ref{eq:Ecore_dipole} shows that the energy of a dislocation
dipole is the sum of a constant term ($E^{\rm core} + E_{\rm c}^{\rm elas}(\phi)$)
and a term depending on the distance $d$ between the two dislocations
composing the dipole. 
The values obtained from atomic simulations and anisotropic linear elasticity
(Tab.~\ref{tab:Ecore_Fe}) shows that the $1/2\left<111\right>\left\{110\right\}$
edge dipole is more stable than the $1/2\left<111\right>\left\{112\right\}$
one, whatever the distance $d$.
This result is strictly valid only when both dislocations share
the same glide plane ($\phi=0$).
We now need to look to the variation of the distant independent term
$E^{\rm core} + E_{\rm c}^{\rm elas}(\phi)$ with the angle $\phi$,
\ie with the orientation of the dipole.

Fig. \ref{fig:Ephi} shows the variation of $E^{\rm core} + E_{\rm c}^{\rm elas}(\phi)$
with the angle $\phi$ for both dislocation. 
Whatever the orientation of the dipole, this energy contribution 
is smaller for the $1/2\left<111\right>\left\{110\right\}$ than
for the $1/2\left<111\right>\left\{112\right\}$ edge dipole.
The $1/2\left<111\right>\left\{110\right\}$ is therefore 
found as the most stable edge dipole.

\section{Conclusions and discussion}

We obtained within the sextic formalism an expression 
of the core traction contribution to the elastic energy (Eq. \ref{eq:energy_dislo_core}).
We showed that this expression agrees with the angular
dependence of the interaction energy between two dislocations 
previously derived by Stroh \cite{STR58}.
This energy contribution is actually important so that the 
dislocation elastic energy can be a state variable 
consistent with the work of the Peach-Koehler forces.

This contribution to the elastic energy needs also to be considered
when one wants to extract from atomic simulations core energies 
which does not depend on the simulation conditions.
Both the core traction contribution and the core energy
do not depend on the length scale of the dislocation microstructure.
Their physical meaning is nevertheless different. 
The core energy is a dislocation intrinsic property which
takes into account the fact that atomic interactions cannot 
be described by elasticity close to the dislocation core.
On the other hand, the core traction contribution is a part
of the elastic energy and depends on the positions of other
dislocations through the angle $\phi$ defining the dislocation cut.

The application to edge dislocations in iron showed that 
dislocation core energies could be obtained consistently 
from different simulation approaches, \ie the cluster or 
the dipole approaches. Both approaches lead to the same core energy
when one does not forget to take into account 
the energy contribution of core tractions in the dipole approach. 
This allowed us to conclude that 
$1/2\left<111\right>\left\{110\right\}$ edge dislocations 
are more stable than $1/2\left<111\right>\left\{112\right\}$
ones.

The contribution of core tractions is also important 
when defining the elastic energy of a dislocation loop
as pointed by Bullough and Foreman \cite{BUL64}.
For a glissile loop, this contribution generally reduces the
elastic energy of edge segments: this is true at least for an isotropic
crystal (Eq.~\ref{eq:Ec_iso_edge}). 
As a consequence, this would lead to loop shapes 
which are more rounded than with a simple line tension model
where this contribution is omitted \cite{WIT59}. 
As this contribution does not depend on the loop size,
smaller loops should be more rounded than larger ones, 
in agreement with experimental observations \cite{MUG01}.
This energy contribution could therefore explain some discrepancies
obtained between a simple line tension model and, either experimental
observations \cite{MUG01}, or results deduced from simulations 
of the loop self stress \cite{SCH88}.

Gavazza and Barnett \cite{GAV76} showed that part of the self-force
acting on a loop is associated with the core traction contribution 
to the elastic energy.
The obtained expression of this contribution to the elastic energy 
(Eq.~\ref{eq:energy_dislo_core}) could therefore be used 
in dislocation dynamics simulations to compute the self-force
acting on a dislocation segment.

To conclude, we would like to stress that we obtained the expression 
of the core traction contribution to the elastic energy 
within the framework of the sextic formalism. 
Anisotropic linear elasticity of line defects can also be handled 
within the integral formalism\cite{BAC80,TIN96}.
In the integral method, all quantities defining the 
elastic fields are obtained from angular integrals,
so that one does not need to look for the solutions
of the sextic equations. 
The recasting of our result
within the integral formalism stills needs to be done. 
But this may not be possible as the expression \ref{eq:energy_dislo_core}
we obtained does not make appear any known integral. 
Getting an expression of the core traction contribution
to the elastic energy within the integral
formalism may be therefore a challenge.

\appendices

\section*{Acknowledgments}
The author is grateful to V. Bulatov, L. Ventelon, and 
F. Willaime for stimulating discussions,
as well as to B. Lemoine for his help with the atomic simulations.


\section{Integral $J^1(p,q)$}
\label{sec:integral_log_J1}

In this appendix, we calculate the integral appearing 
in the contribution to the dislocation elastic energy 
of the core tractions:
\begin{equation*}
	J^1(p,q) = \int_{-\upi}^{\upi}{\frac{-p\cos{\theta}+\sin{\theta}}
	{\cos{\theta}+p\sin{\theta}} 
	\log{\left( \cos{\theta}+q\sin{\theta} \right)}\md{\theta}}.
\end{equation*}

An integration by parts leads to
\begin{multline*}
	J^1(p,q) = - \left[ \log{\left( \cos{\theta}+p\sin{\theta} \right)}
	\log{\left( \cos{\theta}+q\sin{\theta} \right)}	\right]_{-\upi}^{\upi} \\
	+ \int_{-\upi}^{\upi}{\log{\left( \cos{\theta}+p\sin{\theta} \right)}
	\frac{-\sin{\theta}+q\cos{\theta}}
	{\cos{\theta}+q\sin{\theta}} 
	\md{\theta}}.
\end{multline*}
We can conclude that
\begin{equation}
	J^1(p,q) = - J^1(q,p)
	\label{eq:J1_property}
\end{equation}

It is not possible to directly perform the integration,
because of the logarithm function appearing in the definition
of $J^1(p,q)$.
To circumvent the problem, we derive $J^1(p,q)$ with respect to the 
parameter $q$
\begin{equation*}
	\frac{\upartial{J^1(p,q)}}{\upartial{q}}
	= \int_{-\upi}^{\upi}{\frac{-p\cos{\theta}+\sin{\theta}}
	{\cos{\theta}+p\sin{\theta}} 
	\frac{\sin{\theta}}{\cos{\theta}+q\sin{\theta}} \md{\theta}}.
\end{equation*}
We thus obtain an integral of a rational function of $\cos(\theta)$ 
and $\sin(\theta)$. It can be integrated using the residues 
theorem \cite{ARF01}:
\begin{align*}
	\frac{\upartial{J^1(p,q)}}{\upartial{q}} 
	& = -\frac{2\upi\mi}{\mi+q}                       
		& \textrm{ if } \Im{(p)}>0 \textrm{ and } \Im{(q)}>0 &,\\
	& = -\frac{2\upi\mi}{\mi-q} + \frac{4\upi\mi}{p-q} 
		& \textrm{ if } \Im{(p)}>0 \textrm{ and } \Im{(q)}<0 &,\\
	& = -\frac{2\upi\mi}{\mi+q} + \frac{4\upi\mi}{q-p} 
		& \textrm{ if } \Im{(p)}<0 \textrm{ and } \Im{(q)}>0 &,\\
	& = -\frac{2\upi\mi}{\mi-q}                       
		& \textrm{ if } \Im{(p)}<0 \textrm{ and } \Im{(q)}<0 &.
\end{align*}
Now, we can integrate this result with respect to the parameter $q$:
\begin{align*}
	J^1(p,q) 
	=& -2\upi\mi\log{(\mi+q)}                     + f_1(p) 
		& \textrm{ if } \Im{(p)}>0 \textrm{ and } \Im{(q)}>0 &,\\
	=&  2\upi\mi\log{(\mi-q)} - 4\upi\mi\log{(p-q)} + f_2(p)
		& \textrm{ if } \Im{(p)}>0 \textrm{ and } \Im{(q)}<0 &,\\
	=& -2\upi\mi\log{(\mi+q)} + 4\upi\mi\log{(q-p)} + f_3(p) 
		& \textrm{ if } \Im{(p)}<0 \textrm{ and } \Im{(q)}>0 &,\\
	=&  2\upi\mi\log{(\mi-q)}                     + f_4(p) 
		& \textrm{ if } \Im{(p)}<0 \textrm{ and } \Im{(q)}<0 &,
\end{align*}
where $f_1$, $f_2$, $f_3$ and $f_4$ are four functions depending solely
on the parameter $p$. They can be determined by using the property
\ref{eq:J1_property} and the result $J^1(\mi,-\mi)=2\upi^2$.
This leads to the final result:
\begin{align*}
	J^1(p,q) 
	= & 2\upi\mi\log{(\mi+p)} - 2\upi\mi\log{(\mi+q)}
		& \textrm{if } \Im{(p)}>0 \textrm{ and } \Im{(q)}>0 &,\\
	= & 2\upi\mi\log{(\mi+p)} + 2\upi\mi\log{(\mi-q)} \\
	  & - 4\upi\mi\log{(p-q)}
	  	& \textrm{if } \Im{(p)}>0 \textrm{ and } \Im{(q)}<0 &,\\
	= & -2\upi\mi\log{(\mi-p)} -2\upi\mi\log{(\mi+q)} \\
	  &+ 4\upi\mi\log{(q-p)}
	  	& \textrm{if } \Im{(p)}<0 \textrm{ and } \Im{(q)}>0 &,\\
	= & -2\upi\mi\log{(\mi-p)} + 2\upi\mi\log{(\mi-q)}
		& \textrm{if } \Im{(p)}<0 \textrm{ and } \Im{(q)}<0 &.
\end{align*}

\section{Integral $\Delta J_{\phi}^1(p,q)$}
\label{sec:integral_log_dJ1}

The angular dependence of the dipole elastic energy
makes appear the following integral
\begin{multline*}
	\Delta{J^1_{\phi}}(p,q) 
	= \int_{-\upi}^{\upi}
	\frac{-p\cos{\theta}+\sin{\theta}}
	{\cos{\theta}+p\sin{\theta}} \\
	 \left[ \log{\left( \frac{ \cos{\theta}+q\sin{\theta} }
	{\cos{\phi}+q\sin{\phi}}\right)}
	 - \log{\left( \cos{\theta}+q\sin{\theta} \right)}
	\right]\md{\theta}.
\end{multline*}
We use the same integration method as for $J^1\left( p,q \right)$. 
A derivation with respect to the parameter $q$ leads to 
\begin{equation*}
	\frac{\upartial}{\upartial q}{\Delta{J^1_{\phi}}(p,q)}
	= \int_{-\upi}^{\upi} \frac{-p\cos{\theta}+\sin{\theta}}
	{\cos{\theta}+p\sin{\theta}}
	\frac{-\sin{\phi}}{\cos{\phi}+q\sin{\phi}}
	\md{\theta}.
\end{equation*}
Using the residues theorem, we obtain
\begin{align*}
	\frac{\partial}{\partial q}{\Delta{J^1_{\phi}}(p,q)}
	= & 2\upi\mi\frac{\sin{\phi}}{\cos{\phi}+q\sin{\phi}}
	&\textrm{if } \Im{(p)}>0 , \\
	= & -2\upi\mi\frac{\sin{\phi}}{\cos{\phi}+q\sin{\phi}}
	&\textrm{if } \Im{(p)}<0 .
\end{align*}
We then integrate with respect to the parameter $q$
and we use the property $\Delta{J^1_{\phi}}(p,q) 
= - \Delta{J^1_{\phi}}(q,p)$ deduced from an integration
by parts of the initial integral.
This leads to the final result
\begin{align*}
	\Delta{J^1_{\phi}}(p,q)
	=& 2\upi\mi \log{\left[ \cos{\phi}+q\sin{\phi} \right]} \\
	 & -2\upi\mi \log{\left[ \cos{\phi}+p\sin{\phi} \right]} 
	  &\textrm{if } \Im{(p)}>0 \textrm{ and } \Im{(q)}>0 ,\\
	=& 2\upi\mi \log{\left[ \cos{\phi}+q\sin{\phi} \right]} \\
	 & +2\upi\mi \log{\left[ \cos{\phi}+p\sin{\phi} \right]} 
	  &\textrm{if } \Im{(p)}>0 \textrm{ and } \Im{(q)}<0 ,\\
	=& -2\upi\mi \log{\left[ \cos{\phi}+q\sin{\phi} \right]} \\
	 & -2\upi\mi \log{\left[ \cos{\phi}+p\sin{\phi} \right]} 
	  &\textrm{if } \Im{(p)}<0 \textrm{ and } \Im{(q)}>0 ,\\
	=& -2\upi\mi \log{\left[ \cos{\phi}+q\sin{\phi} \right]} \\
	 & +2\upi\mi \log{\left[ \cos{\phi}+p\sin{\phi} \right]} 
	  &\textrm{if } \Im{(p)}<0 \textrm{ and } \Im{(q)}<0 .
\end{align*}


\end{document}